\documentclass[10pt,letterpaper]{article}

\usepackage{amsmath} 
\usepackage{array} 
\usepackage{threeparttable}
\usepackage{float}

\renewenvironment{abstract}
{
  \centerline
  {\large \bfseries \scshape Abstract}
  \begin{quote}
}
{
  \end{quote}
}

\usepackage[top=1in,left=1in]{geometry}

\usepackage[utf8]{inputenc}

\usepackage{cite}

\usepackage{nameref,hyperref}

\usepackage{microtype}
\DisableLigatures[f]{encoding = *, family = * }

\raggedright
\usepackage{changepage}

\usepackage[aboveskip=1pt,labelfont=bf,labelsep=period,singlelinecheck=off]{caption}

\makeatletter
\renewcommand{\@biblabel}[1]{\quad#1.}
\makeatother

\usepackage{color}

\definecolor{Gray}{gray}{.25}

\usepackage{graphicx}

\usepackage{sidecap}

\usepackage{wrapfig}
\usepackage[pscoord]{eso-pic}
\usepackage[fulladjust]{marginnote}
\reversemarginpar

\begin{document}
\vspace*{0.35in}

\begin{flushleft}
{\Large
\textbf\newline{\LaTeX \textbf{\textbf{ DETERMINANTS OF SAVING BEHAVIOR AMONG EMPLOYEES IN DHAKA, BANGLADESH}
}
}
\newline
\\
\centerline{\small{\bfseries Soumita Roy \textsuperscript{1,*} Md Muntasir Kamal Dihan \textsuperscript{1,*} }}


\centerline{\small{\bfseries Tasnimah Haque \textsuperscript{1,*}  Nafisa Nomani\textsuperscript{1,*} Sadia Islam Preety \textsuperscript{1,*}  
}}

}
\centerline{\textsuperscript{1}\small{Institute of Business Administration - JU}}
\end{flushleft}
\bigskip

\begin{abstract}
\textbf{Purpose} –With an emphasis on elements like financial knowledge, financial attitude, social influence, financial self-efficacy, and financial management practices, this study explores the factors that influence employees' saving behavior in Dhaka, Bangladesh.  We also welcome others to work on saving behavior, which is the main reason for publishing. The purpose is to make others aware of the methods for quantitative financial behavior analysis in Bangladesh. 

\textbf{Design/methodology/approach} – The study uses a quantitative approach with a cross-sectional survey design. Data was collected from 40 participants through a structured questionnaire adapted from reliable sources. The questionnaire captured demographic information and used established items to measure the key variables. Data analysis included descriptive statistics, reliability analysis using Cronbach's alpha, and regression analysis to test the hypothesized relationships.

\textbf{Findings }– The results indicate that among the factors examined, only financial management practices had a significant positive relationship with saving behavior. Rest of the factors did not show significant relationships with saving behavior in this study sample. 

\textbf{Limitation / Disclaimer }– It is still a work in progress, this paper is meant for pre-print with mostly incomplete and limited data. No data cleaning was performed, so it is very likely to include outliers and faulty data. Realizing the necessity for the method to determine saving behavior analysis in Bangladesh and novelty of the method, the decision to pre-publish is being realized. We encourage not to use it for any statistical work at the moment, as this research progress required to recollect data, but this paper shows the method we will use to analyze. We also welcome others to work on saving behavior, which is the main reason for publishing.

\textbf{Originality/value} – This study contributes to the limited research on saving behavior determinants in the Bangladeshi context, specifically among employees in the capital city of Dhaka. It explores the influence of multiple factors, including the rarely studied aspect of social influence.

\end{abstract}


\bigskip
\textbf{Keywords} Saving behavior, financial behaviour, financial literacy, financial attitude, social influence, financial management.

\bigskip
\textbf{Paper type} Work In Progress Research Paper

\bigskip
\textbf{
\section{Introduction}
}

This study investigates the factors that determine saving behavior among the employees of Dhaka city in Bangladesh. Saving behavior refers to the habit and practice of keeping aside a portion of one’s current income for the future, instead of spending the entire amount (K. Tharanika and Andrew, 2017). Saving behavior is an important aspect of an individual’s life because it has an impact on both micro and macro levels. On the micro level, saving behavior significantly contributes to the financial well-being of a person (Ingale \& Paluri, 2020). It also collectively influences economic growth on the macro level because when people save, they contribute to the capital accumulation of a country (Jumena et al., 2022). Saving behavior plays a crucial role in fulfilling current needs, ensuring financial stability, and meeting future financial goals. 

Dhaka, being the capital and economic hub of Bangladesh contributes to 36\% of the country’s GDP (Banu, 2023). The city has numerous government and private offices, making it the source of earnings for thousands of people. Despite the importance of saving behavior, there is a noticeable research gap regarding the determinants of saving behavior among employees in the Bangladeshi context, especially in Dhaka city. Existing studies in this area are limited in scope and fail to provide a comprehensive understanding of the factors influencing saving behavior. Consequently, there is a need for deeper research to explore the role of various factors, including financial knowledge, attitude, self-efficacy, social influence, and financial management practices, in shaping saving behavior among employees in Dhaka so that they can get access to better financial well-being, as well as can contribute to country’s economy. 

The study will explore the impact of financial knowledge, financial attitude, financial self-efficacy, social influence, and financial management practices on the saving behavior of employees. Financial knowledge can be defined as the understanding of financial concepts, products, and strategies by an individual (Morris et al., 2022). Financial attitude deals with an individual's beliefs, values, and emotions towards saving and financial management. Financial self-efficacy pertains to an individual's confidence in their ability to effectively managing their finances. Social influence refers to the influence of family, friends, and colleagues on an individual's saving behavior (Ismail et al., 2013). Financial management practices include budgeting, investment decisions, and debt management. The paper will conclude by summarizing the key findings, implications, and recommendations for promoting saving behavior among employees in Dhaka, Bangladesh.

\bigskip
\bigskip
\textbf{
\section{Literature Review}
}

\subsection{ \textit{Financial knowledge and saving behavior}
}

Financial behaviors, like borrowing, saving, and investing decisions made by individuals, are directly associated with the notion of financial knowledge (Lusardi \& Mitchell, 2014). Strong financial literacy empowers people to create detailed financial plans and make well-informed financial decisions, which facilitates efficient personal money management and a grasp of how financial institutions operate (Mahdzan \& Tabiani, 2013).

According to research, financial literacy has a positive correlation with the capacity to plan through better insurance and savings plans, (Hinga, 2014). Furthermore, populations that have reached their forties tend to have higher financial knowledge than younger or older populations, which may impact their capacity to make sound financial decisions (Potrich, Vieira, Kirch, 2015). In contrast, those lacking financial literacy are more likely to seek loans more frequently,  and pay more for financial products (García, Santillán, Tiburcio, 2013). 

Financial awareness varies across demographics, with women and those with lower education levels often underrepresented in financial education courses (García et al., 2013). However, having more financial information does not guarantee better financial management outcomes; rather, it must be linked with cultivated values and appropriate financial behaviours (Qamar, Khemta, \& Jamil, 2016).

Keller and Staelin (1987) stated that there are formal and informal approaches to learning about money. Formal sources include college or high school courses; informal sources include peer advice, training sessions, and seminars. A person's wealth, income, age, and level of education all have an impact on how much money they save. (Browning and Lusardi, 1996). Those with a great understanding of finance are more likely to engage in responsible financial behaviours, whereas those with a limited awareness of the need of saving may make frequent mistakes when managing their money (Hilgert, Hogarth, and Beverly, 2003; Bodie, 2002).

\bigskip

\textbf{\textit{
{H\textsubscript{1}: There is a significantly positive relationship between financial knowledge and saving behavior.}
}}
\bigskip
\bigskip
 \textit{
\subsection{Financial attitude and saving behavior}
}

Financial attitude refers to an individual's perception of money, which has been shown in previous studies to be a significant factor influencing saving behavior (Hrubes, Ajzen, \& Daigle, 2001). Individuals' attitudes can affect how they perceive their money, as well as their social preferences and self-esteem. (Gasiorowska, 2015). These differences in attitudes are shaped by various factors such as age, wealth, personality or social class, which in turn influence one's attitude toward money (Ismail, 2020; Gasiorowska, 2015).

The way people use, save, and occasionally mismanage their money is greatly influenced by the attitudes they have towards money. For example, people with low and moderate incomes may find it difficult to save since they must make ends meet on a daily basis (Hayhoe, Cho, DeVaney, Worthy, Kim \& Gorham, 2012). 

Financial attitude encompasses responses such as liking or disliking financial behaviors, and it shapes spending habits, saving behaviors, and wasteful spending tendencies (Potrich et al., 2016; Furnham, 1984). Financial attitude affects a variety of financial management actions, such as saving, spending, and investing, claim (Mien \& Thao, 2015).

Financial behavior, on the other hand, reflects actions that demonstrate responsible management of financial resources to achieve personal financial objectives (Potrich et al., 2016). It involves individual responsibility in managing finances to meet both needs and desires, as highlighted by Nababan \& Sadalia (2013).

\bigskip
\textbf{\textit{H\textsubscript{2}: There is a significantly positive relationship between financial attitude and saving behavior.}}

\bigskip\bigskip
 \textit{
\subsection{Financial self-efficacy and saving behavior}
}

Self-confidence and financial competence are linked to financial self-efficacy, which is described as a person's capacity to handle their finances (Lapp, 2010). (Lown, 2011).Financial self-efficacy is linked to self-confidence and financial skill (Lown, 2011). It is described as an individual's ability to manage their finances (Lapp, 2010). In a broader sense, self-efficacy is the conviction that one can handle a circumstance successfully (Letkiewicz et al. 2014). In terms of finances, it investigates the psychological inclination that promotes actions that contribute to improved financial well-being and decision-making. Higher self-efficacy people are more assured and adept at handling financial difficulties than lower self-efficacy people (Kraft et al., 2005). 

This notion goes beyond finances, with high self-efficacy likely to result in total well-being, This idea extends beyond money; according to Letkiewicz et al. (2014), having a high level of self-efficacy is likely to lead to overall well-being, which encompasses both mental and physical health as well as the ability to modify behavior. According to Letkiewicz et al. (2014), people who have high financial self-efficacy can handle their money well and ask for help when needed, while people who have low self-efficacy have trouble with both. Furthermore, higher financial self-efficacy has been associated with lower debt, minor financial issues, higher savings, and lower financial stress (Lapp, 2010).

\bigskip

\textbf{\textit{ H\textsubscript{3}: Financial self-efficacy has significant relationships with saving behavior.}
}

\bigskip
\bigskip

\subsection{ \textit{Social Influence and Saving Behaviour}
}

According to Franzoi (2006), social influence is when people or organisations use their social power to persuade others to adopt a particular attitude or set of behaviours. In the framework of this investigation, social influence includes peer and parent socialisation.

Several research works have shown how parents play a critical role in influencing their children's financial socialization (Cude et al., 2006). Parents wield considerable influence in shaping their children's financial behavior and are regarded as primary role models in managing financial affairs. Economic socialization, particularly discussions about financial matters within the family, has been shown to impact children's future orientation positively (Webly \& Nyhus, 2006). Consequently, children with strong family relationships are more inclined towards future-oriented and financially responsible behaviors.

Shim et al. (2010) found that parental roles had a substantial impact on young adults' financial education and work experience compared to other factors. Parents and other family members' supportive social support networks are essential for helping young adults and adolescents grow up to be successful adults. Good financial practices are a great way to set a positive example for your kids and encourage positive attitudes and actions in young adults.

Parenting variables have been found to be significantly linked to credit card problems and debt found in college students, as parenting facilitation emerging as a key influencer on credit card usage (Norvilitis \& Maclean, 2010). Hands-on approaches by parents in teaching money management skills, allowances, and bank account management motivate students to reduce credit card usage during college years. Childhood experiences play a crucial role in shaping individuals' behaviors and attitudes in adulthood, highlighting the critical role of parents in financial socialization.

Beyond parenting factors, peer influence also plays an important role in shaping financial behaviors. Similar arguments have been supported by research indicating that individuals with similar preferences tend to form groups, leading to correlations between group and individual behaviors (Duflo \& Saez, 2001).

Individuals draw motivation from various social sources, including family, friends, and employers. Social environments, particularly familial influences, significantly impact attitudes and behaviors (Clarke, 2005). Friends also exert influence over savings behaviors (Duflo \& Saez, 2003), as peer influence is acknowledged as a factor in determining both positive and negative behaviors, attitudes, and values in teenagers (Cohen, 1983).

\bigskip

\textit{ H\textsubscript{4}: Social Influence has significant relationships with saving behavior.}

\bigskip\bigskip

\subsection{ \textit{Financial management practice and saving behavior}
}

Cost, frequency, and importance must regularly be balanced when making financial decisions. According to Mien and Thao (2015), financial management encompasses a variety of actions related to investing, retirement, insurance, and money management. Good financial managers refrain from overspending and follow suggested procedures within a financial management framework (Pham, Yap, and Dowling, 2012). Cost, frequency, and importance must regularly be balanced when making financial decisions.

Financial management, both individual and household, has four essential components. which include setting aside money for savings and consumption, paying bills, limiting smaller personal expenses, and managing daily spending (Krah et al., 2014). People who regularly follow good financial management practices typically get optimistic about their finances and gain trust in them for the future as well as the present. For instance, these people are more likely to place a higher priority on conserving money for future need than on impulsive purchases made within a tight budget. When financial management principles are applied consistently, people feel more optimistic because they believe they are well-prepared and capable of saving money. Effective financial management also includes carefully distributing income across different time periods so that a steady level of spending can be maintained even if income changes over the course of a person's life.

\bigskip
\textit{ H\textsubscript{5}: There is a positive relationship between financial management and saving behavior. }

\bigskip

\section{ \textbf{Research Gap}
}

 \textit{Limited research in the Bangladeshi context:} While existing research explores the factors influencing savings behavior, there's a dearth of studies specifically examining this phenomenon in Bangladesh. This lack of context-specific knowledge hinders our understanding of how cultural and social factors unique to Bangladesh might influence saving patterns.

\textit{Unexplored role of social influence:} Current research often investigates the impact of several variables on savings behavior. However, the influence of social networks, particularly family and friends, remains largely unexplored in this context. Given the collectivistic nature of Bangladeshi society, where social connections are strong, incorporating this variable presents a novel perspective and potentially significant insights.

\bigskip
 \textbf{
\section{Objectives}
}

\begin{enumerate}
    \item To identify the most significant factor influencing saving behavior among the target population.
    \item To provide insights and recommendations for relevant organizations and policymakers to develop strategies and create targeted financial services for promoting personal savings and avoiding financial distress.
    \item To develop a deeper understanding of financial decision-making processes.

\end{enumerate}

\bigskip
\textbf{
\section{Methodology}
}
\bigskip
 \textit{Research Design}: The respondents were employed working class individuals who reside in Dhaka, Bangladesh. 35 questionnaires were developed for the quantitative approach using a survey design of 5 items for each of the 6 variables (Financial Knowledge, Financial Attitude, Financial Self-Efficacy, Social Influence, Financial Management Practices, Savings Behavior). This survey was employed as the data were gathered at a one-time period. We used Jamovi to record and analyze data. Cronbach's alpha and a reliability test were used to assess the content validity. Finally, each of the hypotheses was tested using regression analysis using Jamovi.

\bigskip

\textit{Data Collection}: A structured questionnaire was developed by adapting reliable and validated instruments from the base paper and other relevant sources. The questionnaire consisted of two sections: demographics and selection determinants of saving behavior.

\bigskip
\textit{Data collection and descriptive analysis}: This study investigated the relationship between financial literacy and saving behavior among employees in Dhaka city. Due to limitations in time, resources, and the unknown nature of the target population, a convenience sampling technique will be used to recruit 40 participants. Efforts were made to ensure a diverse sample by including employees from various organizations, income levels, and demographics.
\bigskip
The data were collected through a validated questionnaire containing two sections:

\textit{Demographic Information}\textbf{:} This section gathered basic information about the participants, including age, gender, occupation, education, household size, and monthly family income.

\bigskip

Demographics information across the following categories were collected:

· Age: The questionnaire captures age groups ranging from below 25 to above 50.

· Gender: It includes options for Male, Female, and Other.

· Occupation: It allows participants to identify as Self-Employed Professionals, Officers/Executives (Junior, Middle/Senior), or Others.

· Education: The options include Graduate/Postgraduate (Professional and General), and Some College but not graduate.

· Monthly Family Income: The questionnaire offers eight income brackets from below 30 thousand to above 10 lakh taka.

\bigskip
\textit{The variables are as follows:}

1. Financial Knowledge: Interest rates, risk, inflation. 

2. Financial Attitude: Saving habits, responsibility, Comparison shopping. 

3. Financial Self-Efficacy: Manage money, solve problems, avoid impulse buys. 

4. Social Influence: Friends, family, role models on saving. 

5. Financial management practices: retirement plan, track expenses, budgets, emergency funds, goals. 

6. Saving habits: Frequency, consistency, fund for emergency, spending awareness.

 \bigskip
 \bigskip 
\section{ \textbf{Findings}
}
 \textit{\textbf{Frequency statistics}}

The survey was completed by forty respondents in total. The sample's demographic data is summed together in Table 1. The sample had varied in a number of ways. There were 45\% (18) female responses and 55\% (22) male respondents.

\bigskip
The respondents' ages were distributed as follows: 27.5\% (11) of the participants were under 25, 25\% (10) were between 25 and 30 years old, 30\% (12) were between 31 and 40 years old, 12.5\% (5) were between 41 and 50 years old, and 5\% (2) were above 50.

\bigskip
Regarding employment, professionals working for themselves made up 35\% (14) of the total, followed by junior officers/executives (22.5\%), middle/senior officers/executives (27.5\%), and others (15\%).

\bigskip
Regarding education, 22 graduates or postgraduates with general degrees made up 55\%, 15 graduates or postgraduates with professional degrees was 38\%, and 8\%(3) had some college education but no degree, and 0\% had other education qualifications.

\bigskip
 \textit{Table 1 Sample Characteristics }

\begin{table}[H]
\centering

\begin{tabular}{l l l l}
\hline
\textbf{Demographics} & \textbf{Factors} & \textbf{Frequency} & \textbf{Percentage} \\
\hline
Age Group & Below 25 & 11 & 27.5\% \\
  & 25-30 & 10 & 25.0\% \\
  & 31-40 & 12 & 30.0\% \\
  & 41-50 & 5 & 12.5\% \\
  & Above 50 & 2 & 5.0\% \\
Gender & Male & 22 & 55.0\% \\
  & Female & 18 & 45.0\% \\
  & Other & 0 & 0.0\% \\
Occupation & Self Employed Professional & 14 & 35.0\% \\
  & Officers/Executives - Junior & 9 & 22.5\% \\
  & Officers/Executives-Middle / Senior & 11 & 27.5\% \\
  & Others & 6 & 15.0\% \\
Education & Graduate/Postgraduate - Professional & 15 & 37.5\% \\
  & Graduate/ Post Graduate General & 22 & 55.0\% \\
  & Some College but not graduate & 3 & 7.5\% \\
  & other  & 0 & 0.0\% \\
Monthly Family Income & Below 30 thousand & 6 & 15.0\% \\
(in BDT) & Below 50 thousand & 7 & 17.5\% \\
  & Below 70 thousand & 5 & 12.5\% \\
  & Below 1 lakh & 10 & 25.0\% \\
  & Below 2 lakh & 8 & 20.0\% \\
  & Below 5 lakh & 2 & 5.0\% \\
  & Below 10 lakh & 2 & 5.0\% \\
  & Above 10 lakh & 0 & 0.0\% \\
\hline

\end{tabular}

\end{table}

 \textit{Data Analysis:}

\bigskip
Jamovi software was used to evaluate the gathered data. Examining the relationship between the independent variables (financial knowledge, attitude, self-efficacy, social impact, and managerial methods) and the dependent variable (saving behavior) is the study's main goal. This made it easier to ascertain how the sample population's saving habits are influenced by various components of financial literacy.

\bigskip
These statements, which include responses on a Likert scale (e.g., Strongly Disagree, Disagree, Neutral, Agree, Strongly Agree), are commonly used in surveys that cover various facets of financial literacy.

\bigskip
 \textit{\textbf{Descriptive Statistics}}

\bigskip
Financial Knowledge: The question with the highest mean score was 'knowing interest rates, financial costs, and borrowing conditions' (3.425), followed by 'understanding financial concepts' (3.275). The lowest scoring item was 'understanding how to minimize risk in investments' (2.9). The standard deviation here ranges from 1.24 to 1.32, depending on the elements.

\bigskip
Financial Attitude: 'comparing prices when purchasing' got the highest mean score (3.65), followed by 'responsible for my financial well-being' (3.30). 'I like to purchase stuff, since it makes me feel good' (3.125) had the lowest score.

\bigskip
Financial Self-Efficacy: The highest mean score was for 'I can figure out a solution when I face financial challenges' (3.375), followed by 'confidence in my abilities to handle finance' (3.375). 'confidence in my ability to manage finance' comes next (3.375). 'sticking to a spending plan if unexpected expenses arise' (2.9) receives the lowest score.

\bigskip
The social influence variable that had the greatest mean (3.2) and STD (1.24) was "My parents/family encourage me to save money regularly." The statement "I follow a role model when making financial decisions" had the lowest mean (2.4) and STD (1.24).

\bigskip
Financial Management Practices: 'Paying all bills on time' (3.375) has the highest mean score, and 'saving for a long-term goal' (3.125) comes in second. 'Creating a monthly budget and trying to stay within that' (2.775) has the lowest score.

\bigskip
Saving Behavior: "Avoiding spending money on unnecessary things" (3.125) has the highest mean score, and "saving money every month' (2.9). The lowest scoring item is 'trying to save in equal amounts' (2.35).

\bigskip
\textit{Table 2 Descriptive Statistics}

\begin{table}[H]
\centering

\begin{tabular}{l l l}
\hline
Variables & Mean & SD \\
\hline
Financial Knowledge & 3.425 & 1.24 \\
Financial Attitude & 3.125 & 1.32 \\
Financial Self-Efficacy & 2.9 & 1.19 \\
Social Influence & 3.2 & 1.24 \\
Financial Management Practices & 3.375 & 1.10 \\
Saving Behavior & 2.9 & 1.32 \\
\hline

\end{tabular}

\end{table}

\bigskip
\textbf{\textit{Reliability Analysis}}

\bigskip
The internal consistency of the study's constructs is measured by reliability. If the Alpha (\$\textbackslash{}alpha\$) value of a construct is higher than.70, it is considered dependable (Hair et al., 2013). Cronbach's Alpha was used to measure construct dependability. All the factors were deemed to be dependable, according to the results. Table 3 provides a summary of the reliability outcomes.

\bigskip
\textit{Table 3 Reliability Statistics}

\begin{table}
\centering

\begin{tabular}{l l l}
\hline
Constructs & No. of items & Alpha(\$\textbackslash{}alpha\$) \\
\hline
Financial Knowledge & 5 & .948 \\
Financial Attitude & 5 & .812 \\
Financial Self-Efficacy & 5 & .784 \\
Social Influence & 5 & .843 \\
Financial Management Practices & 5 & .840 \\
Saving Behavior & 5 & .788 \\
\hline

\end{tabular}

\end{table}
\bigskip
\textbf{ \textit{Assumption checks}
}
The data are found to be normal and homoscedastic. Also, the residuals are not autocorrelated and independent variables are not correlated. (Statistical reports are in Appendix)

\bigskip
\textit{Regression Analysis}

\bigskip
The study seeks to investigate the effect of financial knowledge, financial attitude, financial self-efficacy, social influence and financial management practices on saving behavior. Following hypotheses were proposed.

\bigskip

\textit{\textbf{ H\textsubscript{1}: There is a significantly positive relationship between financial knowledge and saving behavior.}}

\bigskip

\textbf{\textit{H\textsubscript{2}: There is a significantly positive relationship between financial attitude and saving behavior.}}

\bigskip

\textit{\textbf{H\textsubscript{3}: There is a significantly positive relationship between financial self-efficacy and saving behavior.}}

\bigskip

\textbf{\textit{H\textsubscript{4}: There is a significantly positive relationship between social influence and saving behavior.}}

\bigskip

\textbf{\textit{H\textsubscript{5}: There is a significantly positive relationship between financial management practices and saving behavior.}}

\bigskip

Financial knowledge, financial attitude, financial self-efficacy, social influence, and financial management practices were the predictor factors on which the dependent variable (saving behavior) was regressed. F (5, 34) = 20.420, p <.001 shows that the independent variables strongly predict saving behavior, suggesting that the five factors under investigation have a substantial effect on saving behavior. Furthermore, the model explains 75.0\% of the variance in saving behavior, as indicated by the R\textsuperscript{2} = 0.750.

\bigskip
To determine the impact of every factor on the criterion variable (saving behavior), additional coefficient assessments were conducted. H5 assesses whether financial management practices have a major positive impact on saving behavior.  The results revealed that financial management practices have a significant and positive impact on saving behavior (B=.768, t=6.824, p=.000). Hence, H\textsubscript{5} was supported. Surprisingly, the other four factors had a p-value much greater than 0.05. Consequently, H\textsubscript{1}, H\textsubscript{2}, H\textsubscript{3} and H\textsubscript{4} were not supported. 

\bigskip
The results are presented in Table 4.

\bigskip
 \textit{Table 4 Hypotheses results}

\begin{table}[H]
\centering

\begin{tabular}{l l l l l l}
\hline
Hypotheses & Regression Weights & Unstandardized coefficient, B & t & p-value & Results \\
\hline
H\textsubscript{1} & FK → SB & .115 & 1.243 & .222* & Not supported \\
H\textsubscript{2} & FA → SB & -.011 & -.064 & .950* & Not supported \\
H\textsubscript{3} & FS → SB & -.085 & -.529 & .600* & Not supported \\
H\textsubscript{4} & SI → SB & .107 & .978 & .335* & Not supported \\
H\textsubscript{5} & FM → SB & .768 & 6.824 & .000* & Supported \\
\hline
R\textsuperscript{2} & 0.750 &   &   &   &   \\
F (5,34) & 20.420 &   &   & .000* &   \\
\hline

\end{tabular}

\end{table}

\textit{Note.} *p < 0.05. FK: Financial Knowledge, FA: Financial Attitude, FS: Financial Self-Efficacy, SI: Social Influence, FM: Financial Management Practices, SB: Saving Behavior

\bigskip
\textbf{Business Implications }

\bigskip
As the finding suggests there is a significant relationship found between financial management practices and saving behavior. Based on this the potential strategies that can be taken by the businesses and financial institutions are-

\bigskip
\textit{Product Development}: Giving top priority to the creation of resources and tools for financial management that are suited to the requirements and tastes of the intended audience. This could entail developing user-friendly budgeting applications, supplying tailored savings plans, or supplying instructional resources on sensible money management techniques.

\bigskip
\textit{Education and Training} : Putting money into staff training initiatives will guarantee that employees are prepared to offer knowledgeable advice and assistance to clients looking to enhance their money management techniques. To further educate clients on subjects like saving, investing, and budgeting, workshops or seminars may be provided. 

\bigskip
 \textbf{Conclusion and Discussion}

\bigskip
This study investigated the relationship between financial knowledge, attitude, self-efficacy, social influence, and financial management practices with savings behavior among employees of varying ages and positions in Dhaka, Bangladesh. The findings revealed that the combined model including all five factors significantly impacted savings behavior (p-value<0.05). This suggests that the data holds promise for further analysis. The reliability of all the measured variables was confirmed.

\bigskip
However, further analysis showed that only one of the five hypothesized relationships was statistically significant. Hypothesis 5, which proposed a significant association between financial management practices and savings behavior, was supported (p-value<0.05). Financial knowledge, financial attitude, self-efficacy, and social influence did not demonstrate a significant influence on savings behavior in this study.

\bigskip
The inability to confirm the remaining hypotheses could be attributed to the limited sample size. Interestingly, the data suggests a general tendency to save among the Bangladeshi workforce, regardless of their level of financial literacy. This tendency appears to be driven by saving for retirement and emergencies, even if a comprehensive understanding of financial concepts is lacking.

\bigskip
This research offers valuable insights and paves the way for further exploration within the Bangladeshi context. Future studies should prioritize a larger sample size that is more representative of the population. Due to constraints on time, budget, and manpower, this study employed a convenience sampling technique. Future research could potentially generate more robust results by utilizing a more comprehensive sampling method.

\bigskip
The findings presented here hold relevance for financial institutions aiming to enhance customer savings. By interpreting this data, such institutions may gain valuable insights. For example, focusing on encouraging and supporting customer adherence to sound financial management practices could potentially lead to increased savings frequency and amounts.

\bigskip
\bigskip

  \textbf{REFERENCES}
\bigskip

 Banu, N. (2023) ‘Disaster governance and policy for Dhaka: Building a Smarter City’, \textit{International Handbook of Disaster Research}, pp. 1207–1220. doi:10.1007/978-981-19-8388-7\_76. 

Bodie, Z., 2002. Life-cycle Finance in Theory and in Practice. Boston University School of Management Working Paper, (2002-02).

Browning, M. and Lusardi, A., 1996. Household saving: Micro theories and micro facts. \textit{Journal of Economic literature}, \textit{34}(4), pp.1797-1855.

Ismail, S., Koe, W.-L., Mahphoth, M. H., Karim, R. A., Yusof, N., Ismail, S. (2020). Saving behavior determinants in Malaysia: An empirical investigation. KnE Social Sciences, 731–743. https://doi.org/10.18502/kss.v4i6.6639

Cude, B., Lawrence, F., Lyons, A., Metzger, K., LeJeune, E., Marks, L. and Machtmes, K., 2006. College students and financial literacy: What they know and what we need to learn. Proceedings of the Eastern Family Economics and Resource Management Association, 102(9), pp.106-109.

Duflo, E. and Saez, E., 2003. The role of information and social interactions in retirement plan decisions: Evidence from a randomized experiment. The Quarterly journal of economics, 118(3), pp.815-842.

Duflo, E., 2001. Schooling and labor market consequences of school construction in Indonesia: Evidence from an unusual policy experiment. American economic review, 91(4), pp.795-813.

Franzoi, S. L. 2006. Social psychology 4th ed.. New York: McGraw-Hill

Furnham, A., 1984. Many sides of the coin: The psychology of money usage. Personality and individual Differences, 5(5), pp.501-509

Gasiorowska, A., 2015. The impact of money attitudes on the relationship between income and financial satisfaction. Polish Psychological Bulletin, 46(2), pp.197-208.

Gutter, M.S., Hayhoe, C.R., DeVaney, S.A., Kim, J., Bowen, C.F., Cheang, M., Cho, S.H., Evans, D.A., Gorham, E., Lown, J.M. and Mauldin, T., 2012. Exploring the relationship of economic, sociological, and psychological factors to the savings behavior of low‐to moderate‐income households. Family and Consumer Sciences Research Journal, 41(1), pp.86-101.

Hair, J.F., Ringle, C.M. and Sarstedt, M., 2013. Partial least squares structural equation modeling: Rigorous applications, better results and higher acceptance. \textit{Long range planning}, \textit{46}(1-2), pp.1-12.

Hilgert, M.A., Hogarth, J.M. and Beverly, S.G., 2003. Household financial management: The connection between knowledge and behavior. Fed. Res. Bull., 89, p.309.

Hinga, G.K., 2014. \textit{Relationship of financial literacy on individual savings of employees of postal corporation of Kenya based in Nairobi} (Doctoral dissertation, University of Nairobi).

Hrubes, D., Ajzen, I. and Daigle, J., 2001. Predicting hunting intentions and behavior: An application of the theory of planned behavior. Leisure Sciences, 23(3), pp.165-178.

‌Ingale, K.K. and Paluri, R.A. (2020) ‘Financial Literacy and financial behaviour: A Bibliometric analysis’, \textit{Review of Behavioral Finance}, 14(1), pp. 130–154. doi:10.1108/rbf-06-2020-0141. 

Ismail, S., Kamis, R., Hashim, N., Harun, H., \& Khairuddin, N. S. (2013). An Empirical Investigation on Determinants of Attitude towards Saving Behavior. Procedia Economics and Finance, 00, 000-000

Jonubi, A. and Abad, S., 2013. The impact of financial literacy on individual saving: An exploratory study in the Malaysian context. \textit{Transformations in Business \& economics}, \textit{12}(1), p.28.

Jumena, B.B., Siaila, S. and Widokarti, J.R. (2022) ‘Saving behaviour: Factors that affect saving decisions (systematic literature review approach)’, \textit{Jurnal Economic Resource}, 5(2), pp. 217–235. doi:10.57178/jer.v5i2.365. 

K. Tharanika and Andrew, A. (2017). Factors Influencing On Saving Behaviour Among University Students. \textit{International Journal of Research}, 4(14), pp.861–871.

Keller, K.L. and Staelin, R., 1987. Effects of quality and quantity of information on decision effectiveness. \textit{Journal of consumer research}, \textit{14}(2), pp.200-213.

Krah, R.Y., Aveh, F.K. and Addo, R., 2014. An exploratory study of financial management practices among Ghanaian households. \textit{International Journal of Management and Sustainability}, \textit{3}(7), pp.393-414.

Lapp, W.M., 2010. The missing link: Financial self-efficacy’s critical role in financial capability. \textit{San Francisco, CA: EARN Research Institute}, pp.54-63.

Letkiewicz, J.C., Domian, D.L., Robinson, C. and Uborceva, N., 2014. Self-efficacy, financial stress, and the decision to seek professional financial planning help. \textit{Academy of Financial Services}, pp.1-30.

Lim, H., Heckman, S., Montalto, C.P. and Letkiewicz, J., 2014. Financial stress, self-efficacy, and financial help-seeking behavior of college students. Journal of Financial Counseling and Planning, 25(2), pp.148-160.

Lown, J.M., 2011. Development and validation of a financial self-efficacy scale. \textit{Journal of Financial Counseling and Planning}, \textit{22}(2), p.54.

Lusardi, A. and Mitchell, O.S., 2014. The economic importance of financial literacy: Theory and evidence. \textit{American Economic Journal: Journal of Economic Literature}, \textit{52}(1), pp.5-44.

Mahdzan, N. S., \& Tabiani, S. (2013). The Impact of Financial Literacy on Individual Savings. An Exploratory Study in the Malaysian Context. \textit{Transformations in Business and Economics}, 12, 41-55.

Mien, N.T.N. and Thao, T.P., 2015, July. Factors affecting personal financial management behaviors: Evidence from Vietnam. In \textit{Proceedings of the Second Asia-Pacific Conference on Global Business, Economics, Finance and Social Sciences (AP15Vietnam Conference)} (Vol. 10, No. 5, pp. 1-16).

Moreno-García, E., García-Santillán, A. and Munguía-Tiburcio, J.P., 2013. Some aspects about financial knowledge of undergraduate students. \textit{International Journal of Management and Sustainability}, \textit{2}(4), pp.40-62.

Morris, T., Maillet, S. and Koffi, V. (2022) ‘Financial knowledge, financial confidence and learning capacity on financial behavior: A Canadian study’, \textit{Cogent Social Sciences}, 8(1). doi:10.1080/23311886.2021.1996919. 

Nababan, DarmandanSadalia, Isfenti.2013. Analysis of Personal Financial Literacy and Financial Behavior, Strata I Collage studentss, Faculty of Economics, North Sumatra University. Medan: North Sumatra.

Norvilitis, J.M. and MacLean, M.G., 2010. The role of parents in college students’ financial behaviors and attitudes. Journal of economic psychology, 31(1), pp.55-63.

Pham, T.H., Yap, K. and Dowling, N.A., 2012. The impact of financial management practices and financial attitudes on the relationship between materialism and compulsive buying. \textit{Journal of economic psychology}, \textit{33}(3), pp.461-470.

Potrich, A.C.G., Vieira, K.M. and Kirch, G., 2015. Determinants of financial literacy: Analysis of the influence of socioeconomic and demographic variables. \textit{Revista Contabilidade \& Finanças}, \textit{26}, pp.362-377.

Potrich, A.C.G., Vieira, K.M. and Mendes-Da-Silva, W., 2016. Development of a financial literacy model for university students. Management Research Review, 39(3), pp.356-376.

Qamar, M.A.J., Khemta, M.A.N. and Jamil, H., 2016. How knowledge and financial self-efficacy moderate the relationship between money attitudes and personal financial management behavior. \textit{European Online Journal of Natural and Social Sciences}, \textit{5}(2), p.296.

Shim, S., Barber, B.L., Card, N.A., Xiao, J.J. and Serido, J., 2010. Financial socialization of first-year college students: The roles of parents, work, and education. Journal of youth and adolescence, 39, pp.1457-1470. 

Webley, P. and Nyhus, E.K., 2006. Parents’ influence on children’s future orientation and saving. \textit{Journal of economic psychology}, \textit{27}(1), pp.140-164.

\newpage
\textbf{APPENDIX}

\begin{table}[!htb]
\centering
\begin{tabular}{>{\raggedright\arraybackslash}p{0.2\linewidth}>{\raggedright\arraybackslash}p{100mm}l l}

\textbf{Variables} & \textbf{Items} & \textbf{Mean} & \textbf{STD} \\

 {Financial Knowledge}& I understand interest rates, financial costs, and borrowing terms. & 3.425 & 1.24 \\
 & I understand financial concepts such as money management, investment, and budgeting. & 3.275 & 1.22 \\
 & I understand investment returns and risks. & 3.175 & 1.32 \\
 & I understand how to minimize risk in investments. & 2.9 & 1.32 \\
 & Purchasing power decreases when inflation increases & 3.775 & 1.23 \\
 & & &\\
 {Financial Attitude}& I like to buy things, because it makes me feel good. & 3.125 & 1.32 \\
 & I invest regularly & 2.425 & 1.36 \\
 & I am responsible for my financial well-being & 3.3 & 1.07 \\
 & I compare prices when purchasing a product or service. & 3.65 & 1.14 \\
 & I plan in advance how to manage money & 3.1 & 1.17 \\
 & & &\\
 {Financial Self-Efficacy}& I can stick to a spending plan if unexpected expenses arise. & 2.9 & 1.19 \\
 & I do not always borrow money when unexpected expenses occur. & 3.35 & 1.14 \\
 & I can figure out a solution when I face financial challenges. & 3.375 & 1.03 \\
 & I have confidence in my ability to manage finance. & 3.375 & 1.08 \\
 & I do not worry about running out of money after retirement. & 3 & 1.22 \\
 & & &\\
 {Social Influence}& My parents/family encourage me to save money regularly. & 3.2 & 1.24 \\
 & My friends have influenced my attitude towards saving money. & 2.95 & 1.30 \\
 & Seeing people around me facing financial difficulties encourages me to save more. & 3.2 & 1.26 \\
 & The people I interact with regularly promote the importance of saving money. & 3.05 & 1.06 \\
 & I have a role model whom I follow while making financial decisions. & 2.4 & 1.24 \\
 & & &\\
 {Financial Management Practices}& I usually pay all of my bills on time. & 3.375 & 1.10 \\
 & I keep a written or electronic record of my monthly expenses. & 2.75 & 1.30 \\
 & I save for a long-term goal such as a car, education, and home. & 3.125 & 1.16 \\
 & I create a monthly budget and try to stay within that. & 2.775 & 1.14 \\
 & I maintain a fund for emergencies. & 2.95 & 1.20 \\
 & & &\\
 {Saving Behavior}& I save money every month. & 2.9 & 1.32 \\
 & I try to save in equal amounts. & 2.35 & 1.12 \\
 & I record my savings and spending. & 2.675 & 1.16 \\
 & I am aware of how much and where I spend. & 2.85 & 1.19 \\
 & I avoid spending money on unnecessary things. & 3.125 & 1.30 \\ 

\end{tabular}

\end{table}

\begin{table}[!htbp]
\centering

\begin{tabular}{l l l l l l}
\multicolumn{6}{l}{Normality Tests} \\
\hline
\multicolumn{2}{l}{} & \multicolumn{2}{l}{\textbf{Statistic}} & \multicolumn{2}{l}{\textbf{p}} \\
\hline

Shapiro-Wilk &  & 0.952 &  & 0.089 &  \\
Kolmogorov-Smirnov &  & 0.124 &  & 0.527 &  \\
Anderson-Darling &  & 0.879 &  & 0.022 &  \\
\hline
\multicolumn{6}{l}{} \\

\end{tabular}

 \begin{tablenotes}
            \item[1] 
            \raggedright 
H\textsubscript{0}: The data is normal.

H\textsubscript{1}: The data is non-normal.
  \item[2] 
As p-value of Shapiro-Wilk test is greater than 0.05, null hypothesis is accepted. 

That means, the data is normal.

  \item[2]   \item[2] 

        \end{tablenotes}

\end{table}

\begin{table}[!htbp]
\centering
\begin{tabular}{l l l l l l}
\multicolumn{6}{l}{ Heteroskedasticity Tests} \\
\hline
\multicolumn{2}{l}{} & \multicolumn{2}{l}{\textbf{Statistic}} & \multicolumn{2}{l}{\textbf{p}} \\
\hline

Breusch-Pagan &  & 7.17 &  & 0.209 &  \\
Goldfeld-Quandt &  & 1.30 &  & 0.313 &  \\
Harrison-McCabe &  & 0.458 &  & 0.391 &  \\
\hline

\end{tabular}
 \begin{tablenotes}
            \item[1] 
            \raggedright 
H\textsubscript{0}: The data is homoscedastic.

H\textsubscript{1}: The data is not homoscedastic.
  \item[2]
As p-value of Breusch-Pagan test is greater than 0.05, null hypothesis is accepted. 
That means, the data is homoscedastic which means the variance of residuals is constant.

  \item[2]   \item[2] 

        \end{tablenotes}
\end{table}

\begin{table}[!htb]
\centering

\begin{tabular}{l l l l l l}
\multicolumn{6}{l}{Durbin–Watson Test for Autocorrelation} \\
\hline
\multicolumn{2}{l}{\textbf{Autocorrelation}} & \multicolumn{2}{l}{\textbf{DW Statistic}} & \multicolumn{2}{l}{\textbf{p}} \\
\hline

-0.0310 &  & 2.06 &  & 0.950 &  \\
\hline
\multicolumn{6}{l}{} \\
 &  &  &  &  &  \\

\end{tabular}
 \begin{tablenotes}
            \item[1] 
            \raggedright 
H\textsubscript{0}: The residuals are not correlated.

H\textsubscript{1}: The residuals are correlated.

  \item[2]
The number of independent variables is 5 and the sample size is 40. So, the corresponding D\textsubscript{L} and D\textsubscript{U} values are 1.230 and 1.786 at 5\% significance level. Hence, 4- D\textsubscript{U }= 2.214 and 4- D\textsubscript{L }= 2.770. Here, DW statistic is 2.06 which lies in the acceptance region. So, at 5\% significance level, it can be concluded that the residuals are not correlated.

  \item[2]
Also, as p-value of Durbin-Watson test is greater than 0.05, null hypothesis is accepted. That means, residuals are not correlated.

  \item[2]   \item[2] 

        \end{tablenotes}
\end{table}

\begin{table}[H]
\centering

\begin{tabular}{l l l l l l}
\multicolumn{6}{l}{Collinearity Statistics} \\
\hline
\multicolumn{2}{l}{} & \multicolumn{2}{l}{\textbf{VIF}} & \multicolumn{2}{l}{\textbf{Tolerance}} \\
\hline

FK &  & 1.90 &  & 0.527 &  \\
FA &  & 3.91 &  & 0.256 &  \\
FS &  & 2.98 &  & 0.335 &  \\
SI &  & 1.86 &  & 0.536 &  \\
FM &  & 1.81 &  & 0.552 &  \\

\end{tabular}
 \begin{tablenotes}
            \item[1] 
            \raggedright 
H\textsubscript{0}: The independent variables are not correlated.

H\textsubscript{1}: The independent variables are correlated.
  \item[2]
  
VIF value greater than or equal to 4 indicates there is high correlation. Here, collinearity statistics show that none of the independent variables crossed the cut-off point even though Financial Attitude was quite close to the value of 4.

        \end{tablenotes}
\end{table}

\end{document}